\documentclass{article}

\RequirePackage{fix-cm}
\usepackage{graphicx}
\usepackage{amsfonts}
\usepackage{eucal}

\usepackage{amsmath}
\usepackage{amsfonts}
\usepackage{amssymb}
\newcommand{\bea}{\begin{eqnarray}}
\newcommand{\eea}{\end{eqnarray}}
\newcommand{\be}{\begin{equation}}
\newcommand{\ee}{\end{equation}}

\newcommand{\ash}{\textrm{{\rm arcsinh}}}

\newcommand{\sn}{\textrm{{\rm sinn}}}
\newcommand{\nnc}{\nonumber  \\
\nonumber  \\}
\newcommand{\nn}{\nonumber}
\newcommand{\nd}{\vskip .3 cm \noindent}

\graphicspath{%
    {converted_graphics/}% inserted by PCTeX
    {/}% inserted by PCTeX
}
\begin{document}

\title{The Cosmological Constant Constrained with Union2.1 Supernovae Type Ia Data.\\
\vskip2mm
 \large Derivation and evaluation of several FRW and Carmeli models presenting underwhelming support for the {\it standard model}}         % Enter your title between curly braces
\author{Ahmet M. \"O\/ztas$^1$        \and
        Michael L. Smith$^2$}        % Enter your name between curly braces
\date{}          % Enter your date or \today between curly braces
\maketitle

\section*{Abstract}       % Enter section title between curly braces

We derive several, detailed relationships in terms of the Friedmann-Robertson-Walker (FRW) generalization which describe the Universe during both the radiation and matter dominated epochs. We explicitly provide for the influence of radiation, rather than burying this term within the matter term. Several models allow the cosmological constant (CC) to vary with universe expansion in differing manners. We evaluate these and other popular models including the $\Lambda$CDM({\it standard model}), quintessence as presented by Vishwakarma, Equation of State (EoS) and the Carmeli model with data from the 580 Union2.1 supernovae type Ia collection, using several minimization routines and find models built about the CC, the $\Lambda$CDM models, fare no better than those without.
\vskip2mm
\noindent {\it keywords} cosmology -- supernova -- radiation -- analytical methods -- dark energy --  dark matter -- cosmological constant % Enter subsection title between curly braces
\vskip2mm
\noindent{\small $^1$Engineering Physics, Hacettepe University, TR-06800 Ankara, Turkey           
              {\it oztas@hacettepe.edu.tr}}      
\vskip2mm          
\noindent{\small $^2$4S Fuel Research, Inc., Mesa, AZ 85206 USA
	{\it mlsmith55@gmail.com}} \\

\section{Introduction}
\label{intro}
A number of models describing our Universe have been published; many as variations of the Friedmann-Robertson-Walker (FRW) metric presuming a homogeneous and isotropic Universe, of galactic groups as dust particles usually rapidly receding from one another. Many models and astronomical reports designate a new variety of matter, cold dark matter (CDM), displaying properties significantly different from the ordinary\cite{Sumner2012}. Some models and reports also propose a large dependence on a newly hypothesized form of energy, Dark Energy (DE) which is usually and compactly expressed as the cosmological constant (CC)\cite{Carroll2012}. Some new physics proposing modified forms of General Relativity explaining recent astronomical results have also been reviewed\cite{Will2012} and the references covering many models have been recently assembled\cite{Shen2013,Sola2013}.

Two independent groups announced late last century that our Universe is suffering accelerating expansion reporting with data collected from less than 40 supernovae type Ia (SNe Ia) events\cite{Riess1998,Perlmutter1999}. These groups have continued to collect more and better data with the release of the largest collection heretofore of 580 SNe Ia observations, the Supernova Cosmology Project (SCP) Union2.1 compilation; data more accurate and precise than those of a decade ago\cite{Suzuki2012,Astier2012}. This large data collection, culled from many hundreds of observations, may allow us to discriminate between a Universe with or without CDM and suffering DE accelerating expansion, or not, and with flat or curved spacetime geometries and combinations of these situations. 

SN Ia signals are thought to be a rather uniform emissions with characteristics allowing straightforward use as standard candles. It is hoped that supernovae exhibit similar intrinsic luminosities if the emissions originate from type Ia events. Unfortunately, there are many factors involved in supernova light production, dispersal and observations. For many reasons SNe Ia information suffers significant intrinsic errors and each event must be thoroughly analyzed before use as a distance indicator\cite{Calder2013,Wang2012}. The situation is now known to be quite complicated with significant systematic uncertainties\cite{Kirshner2009} and flux-averaging SNe observations has a significant impact on the calculated distance-redshift values\cite{WangWang2013}. The nature of SNe Ia emissions may have changed over the last several billion years which means another correction should be made on top of many other adjustments\cite{WangWang2013}. In addition, intervening dust can modify the emitted light during the long travel\cite{Chotard2011}.

We examine the current {\it standard model} and other, more detailed, models of the FRW universe by allowing differing influences of radiation and matter during various epochs. For some models we presume radiation was an important influence before recombination by the mechanism of elastic collision, and this influence declined drastically at recombination, when atoms were first formed. We presume the influence of radiation has since slackened presently being quite small. For another model we presume DE being synthesized at singularity as an incredibly large though fixed quantity and much like matter and radiation declining in density with increasing Universe volume. We examine a previously introduced model by Vishwakarma where the CC is proportional to the square of the Hubble constant\cite{Vishwa2001}. We also investigate the quintessence model of Peebles and Ratra with a time-dependent CC\cite{Peebles2003,Pavluchenko2003} in terms suggested by Viswakarma\cite{Vishwa2001}. We have previously derived relationships for the FRW with inclusion of polytropic matter, that is, relativistic and non-relativistic matter and have shown the influence of relativistic matter during the current epoch to be quite small as expected\cite{Oztas2011} and do not report such here. Finally, we look at a model of evolving light frequency over time, which is similar in many respects to models considering the slow evolution of SNe Ia explosions over time\cite{SmithOztas2006}. This should also model in a manner similar to emission radiation correlated with host galaxy mass\cite{Kelly2010}. 

We evaluate these models, using both the SNe Ia data as astronomical distance {\em vs.} redshift and {\em vs.} frequency shift. Both methods rely on SNe Ia luminosity distance, with the latter method proportional to the expansion factor and has also been considered by others\cite{BH2011}. Both different analytical techniques correctly weigh errors associated with distance measurements for curve fitting purposes rather than the logarithmic proportionality of more typical treatments (used by many) which wrongly discount the large errors associated with more distant objects. The reader can wonder at the similar sized observational errors reported from signals of SNe Ia at 0.1 and 1.0 redshift separated by billions of years\cite{Kirshner2009}. Our analyses allow inclusion of our location as the abscissa intercept, not usually done, but which we know with certainty.

We examine several popular models which combine the effects of normal, baryonic matter (NM) with CDM into the $\Omega_{m}$ parameter to explain the SNe Ia data. We also examine the simple FRW model, commonly termed the {\it standard model} or the $\Lambda$CDM model, with and without the cosmological constant, $\Omega_{\Lambda}$, with and without spacetime curvature, $\Omega_k$, with and without the equation of state parameter $w$, and with a separate term for the influence of radiation, $\Omega_r$. We also investigate a derivation of the Carmeli General Relativity (CGR) which is explicit only for baryonic matter ($\Omega_b$)\cite{Hartnett2008}.

In general, we find these several models can be discriminated into either good fits or poor, being dependent on the minimization routine selected for the computerized fitting. Those models which include the CC sometimes fit more poorly than those models without. We also find the parameter for normalized matter density to often be much smaller than previously published and more in line with that expected from Big Bang Nucleosynthesis (BBN) studies\cite{Burles1,Burles2}. As judged by application of the Bayesian Information Criterion (BIC) and normalized $\chi^2$ as the goodness of fit, we find the {\em standard model} does not necessarily fit the Union2.1 data collection as well as many other models. Because of the exhaustive analysis most of our results have been placed in the Appendix.

\section{The time dependent cosmological constant}
\label{sec:1}
\nd We begin with two equations of state describing our Universe which are the usual beginnings assuming the Friedmann-Robertson-Walker(FRW) approximations of homogeneous, isotropic matter, energy distribution and metric as
\be
(\frac{\dot {a}}{a})^{2} = \frac{8\pi G}{3}\rho  + \frac{\Lambda(t) }{3} - \frac{k}{R_0^2a^{2}}
\label{EoS1}
\ee
\be\frac{\ddot {a}}{a} = -\frac{4\pi G}{3}(\rho  + 3p ) + \frac{\Lambda(t) }{3}.
\label{EoS2}
\ee
\nd Here $\rho$ is the matter (common and CDM) and energy density, $p$ the pressure, $a$ the expansion factor, G the gravitational constant, $\Lambda(t)$ the CC which is commonly thought to represent the energy responsible for our expanding universe\cite{Riess1998,Perlmutter1999} and $k$ the constant of integration which is usually taken to represent spacetime curvature. We shall argue later that $k$ represents more than simple curvature.

We will allow a time dependent CC, that is,  $\Lambda(t)=\Lambda_{(t)}$ and examine the effect of this variability on the density and expansion parameter relationships. We eliminate $\ddot{a}$ between Eqs.(\ref{EoS1},\ref{EoS2}) and rewrite the FRW system as

\be
\dot {a}^{2} = \frac{8\pi G}{3}\rho a^2 + \frac{\Lambda(t) }{3}a^2 - \frac{k}{R_0^2}
\label{frm3}
\ee
 taking the derivative of this we find
\be
2\dot{a}\ddot{a}=\frac{8\pi G}{3}\dot{\rho}a^2+ \frac{8\pi G}{3}\rho 2 a \dot{a} +  \frac{\dot \Lambda(t) }{3}  a^2+\frac{\Lambda(t) }{3} 2 a \dot{a} 
\ee
\nd then divide each side by $2 a \dot{a}$
\be\frac{\ddot {a}}{a} = \frac{4\pi G}{3}\dot{\rho}\frac{a}{\dot {a}}+\frac{8\pi G}{3}{\rho}+\frac{\dot \Lambda(t) }{6} \frac{a}{\dot {a}}+\frac{\Lambda(t) }{3}.
\label{frm5}
\ee
\nd We now substitute the left hand side of Eq.(\ref{frm5}) with Eq. (\ref{EoS2})
\be
-\frac{4\pi G}{3}\rho  -\frac{4\pi G}{3} 3p  + \frac{\Lambda(t) }{3}=\frac{4\pi G}{3}\dot{\rho}\frac{a}{\dot {a}}+\frac{8\pi G}{3}{\rho}+\frac{\dot \Lambda(t) }{6} \frac{a}{\dot {a}}+\frac{\Lambda(t) }{3}
\ee

\nd after elimination of terms we arrive at a familiar relationship between density and pressure except the additional term of the CC derivative as

\be
\dot{\rho}=-3\frac{\dot {a}}{a}(\rho+p)-\frac{\dot \Lambda(t) }{8\pi G} 
\label{rho_der}
\ee

\nd We now insert the equation of state, $p=w\rho$, as a power law to aid the solution of Eq (\ref{rho_der}) with the details beginning with Eqs. (\ref{rhotint},\ref{soldif}) below.

\be
\rho=a(t)^{-3(1+w)}\left(-\int a(t)^{3(1+w)}\frac{\dot \Lambda(t) }{8\pi G}dt+C\right)
\label{rho_t}
\ee
\nd where $C$ is a constant of integration.

\section{Expansion Parameter Dependency of the Cosmological Constant}
 We propose the evolution of the CC might be a property of the expanding Universe making this dependent on the expansion parameter. The relationship for this evolution is given as

\be
\Lambda(t)=\Lambda(a(t))=\frac{\Lambda(t_0)}{a(t)^m}
\label{Lambda3}
\ee

\nd where $t_0$ is our present time and $m$ is a real number reflecting the state of the Universe. 
\subsection{Cosmological constant dependence with universe volume}
In this section we shall consider DE being similar to normal matter and energy by presuming the quantity of DE has been constant since immediately after singularity. We associate a fixed quantity of DE with the CC, varying in density with the increasing volume of the expanding Universe. One would predict it likely the current DE density would be much smaller than during earlier epochs. This situation does not violate the first law of thermodynamics (by continually creating energy from nothing) as would the situation of spontaneous net DE arising from the vacuum, with all the problems of matching the expectations of quantum mechanics\cite{Carroll2012}. This latter situation of continual synthesis of DE is required by the {\it standard model} which imparts a special property on spacetime contrary to the thoughts of Einstein\cite{Einstein}.

Here we consider a functional structure for the CC and DE as a type of energy complying with the first law of thermodynamics; decreasing within a unit volume as a function of Universe volume increase. In other words, the total quantity of DE remains constant throughout all time. Beginning with Eq. (\ref{Lambda3}) we investigate a positive value for $m$ with $\Lambda(t_0)/a(t)^3$; $m=+3$ for our 3-space.

For example, if we take the CC to be dependent on the expanding volume of the Universe which itself is a function of time as $$ \Lambda(t)\times Volume=constant$$ where $m=3$ for the 3-dimensional (3D) space dependence of the CC. We apply this 3D dependence of the CC to both situations of the radiation and the matter dominated epochs because DE is supposedly a property of spacetime and not a special form of radiation. With this condition Eq.(\ref{rho_t}) can be rewritten as
\bea
\rho&=&a(t)^{-3(1+w)}\left(m\frac{\Lambda(t_0) }{8\pi G}\int a(t)^{3(1+w)}\frac{\dot a(t) }{a(t)^{m+1}}dt+C\right) \nnc
&=&a(t)^{-3(1+w)}\left(m\frac{\Lambda(t_0) }{8\pi G} \int a(t)^{3(1+w)-m-1}\frac{d a(t) }{dt}dt+C\right)\nn.
\eea

\nd We transform this integral by integration across the expansion factor $a$ as
\be
\rho=a(t)^{-3(1+w)}\left(3\frac{\Lambda(t_0) }{8\pi G}\int a^{3(1+w)-m-1}da+C\right). 
\label{rhotint}
\ee 

\subsection{Radiation dominated epoch with decreasing dark energy}
We now consider the radiation dominated epoch as the time after the nucleosynthesis of H isotopes, He and Li (even though the matter concentration was obviously slightly larger than present) and before atomic combination with $\rho=\rho_r$ and  $w=1/3$. With $p=\rho/3 $ as the ideal gas law
we will evaluate Eq. (\ref{rhotint}) for the radiation dominated epoch
\bea
\rho_r&=&a(t)^{-3(1+1/3)}\left(3\frac{\Lambda(t_0) }{8\pi G}\int a^{3(1+1/3)-3-1}da+C\right) \nnc
&=&a(t)^{-4}\left(3\frac{\Lambda(t_0) }{8\pi G}\int da+C\right)
\nnc &=&a(t)^{-4}\left(3\frac{\Lambda(t_0) }{8\pi G}a+C\right) \nnc
&=&3\frac{\Lambda(t_0) }{8\pi G}\frac{1}{a^3}+\frac{C}{a(t)^{4}}. \nn
\label{soldif}
\eea
\nd We solve the constant of integration, $C$, by setting $a(t)=1$ as the present value for both the expansion parameter and radiation density, $\rho_{0,r}$, and
\bea
\rho_{0,r}&=&3\frac{\Lambda(t_0) }{8\pi G}\frac{1}{1^3}+\frac{C}{1^{4}} \nnc
C&=&\rho_{0,r}-3\frac{\Lambda(t_0) }{8\pi G}
\eea
and
\bea
\rho_r&=&3\frac{\Lambda(t_0) }{8\pi G}\frac{1}{a^3}+\frac{C}{a(t)^{4}} \nnc
&=&3\frac{\Lambda(t_0) }{8\pi G}\frac{1}{a^3}+\frac{1}{a(t)^{4}} \left(\rho_{0,r}-3\frac{\Lambda(t_0) }{8\pi G}\right)\nnc
\rho_r&=&\frac{\rho_{0,r}}{a(t)^{4}}+3\frac{\Lambda(t_0) }{8\pi G}\left(\frac{1}{a(t)^{3}}-\frac{1}{a(t)^{4}}\right).
\eea
\nd We now introduce the well-known normalized parameters of 
\be 1 = \Omega_m + \Omega_\Lambda + \Omega_k \label{norm}\ee
\nd where $\Omega_m$ is associated with all types of matter and radiation energy, $\Omega_\Lambda$ is associated with the CC and $\Omega_k$ is the parameter for spacetime curvature with $\Omega_k$ becoming 0 for flat spacetime, and the Hubble constant, $H$, to allow evaluation with astronomical observations
\be
\frac{8\pi G}{3}\rho_r=H_0^2\frac{\Omega_r}{a^4}+3H_0^2\Omega_\Lambda\left(\frac{1}{a^3}-\frac{1}{a^4}\right)
\nn
\ee
\nd or as
\be
\frac{8\pi G}{3H_0^2}\rho_r=\frac{\Omega_r}{a^4}+3\Omega_\Lambda\left(\frac{1}{a^3}-\frac{1}{a^4}\right).
\label{rhotintrad}
\ee
\noindent We see that the evolution of radiation density is both 4D dependent as expected but bears some third order dependency as expected for matter influence.

\subsection{Matter dominated epoch with a decreasing cosmological constant}
To properly evaluate this situation while adhering to energy conservation we shall separate the matter and radiation dominated epochs. We first consider the current matter dominated epoch as containing only pressureless matter.
For evaluation under this condition with $\rho=\rho_m$ and  $w=0$ for pressureless matter, we return to Eq. (\ref{rhotint})
\bea
\rho_m&=&a^{-3(1+0)}\left(3\frac{\Lambda(t_0) }{8\pi G}\int a^{3(1+0)-3-1}da+\rho_{m,0}\right) \nnc
&=&a^{-3}\left(3\frac{\Lambda(t_0) }{8\pi G}\int a^{-1}da+C \right)
\nnc &=&a^{-3}\left(3\frac{\Lambda(t_0) }{8\pi G}\ln(a)+C\right) \nnc
&=&\frac{3}{8\pi G}\frac{\Lambda(t_0)}{a^3}\ln(a)+\frac{C}{a^3}. \nn 
\eea

\nd We solve the constant of integration, $C$, by setting $a(t)=1$ as the present value for both the expansion parameter and matter density, $\rho_{0,m}$ and
\bea
\rho_{0,m}&=&\frac{3}{8\pi G}\frac{\Lambda(t_0)}{1^3}\ln(1)+\frac{C}{1^3}\nnc
C&=&\rho_{0,m},
\eea

\nd and find that
\be
\rho_m=\frac{\rho_{0,m}}{a^3}\frac{3}{8\pi G}+\frac{\Lambda(t_0)}{a^3}\ln(a).
\ee
To associate this situation with observations of SNe Ia emission distances, $D_L$, we presume the typical conditions for normalization of the FRW parameters as above. Using these normalized parameters and the Hubble constant, $H_0$, we rewrite the previous two relationships as

\be
\frac{8\pi G}{3}\rho_m=H_0^2\frac{\Omega_m}{a^3}+3H_0^2\frac{\Omega_\Lambda}{a^3}\ln(a) \nn
\ee
\nd or as

\be
\frac{8\pi G}{3H_0^2}\rho_m=\frac{\Omega_m}{a^3}+3\frac{\Omega_\Lambda}{a^3}\ln (a).
\label{rhotintmat}
\ee

\noindent The matter density is third order dependent as expected, but the influence of DE is not that straightforward.

We now present our derivation to arrive at the useful variable, $D_L$. Because of small but significant differences from the more common derivation we use a form of Eq.(\ref{frm3}) to eventually solve for both the matter and radiation dominated epochs.

\bea
\frac{da}{dt}&=& a\sqrt{\frac{8\pi G}{3}\rho_{m/r}  + \frac{\Lambda(t) }{3} -\frac{k}{R_0^2a^2}} \nnc
da&=&a\sqrt{\frac{8\pi G}{3}\rho_{m/r}  + \frac{\Lambda(t) }{3} -\frac{k}{R_0^2a^2 }} \quad dt \nnc
dt&=&\frac{da}{\displaystyle a \sqrt{\frac{8\pi G}{3}\rho_{m/r}  + \frac{\Lambda(t) }{3} -\frac{k}{R_0^2a^2}}}.
\label{dt}
\eea
\nd We use the null geodesic of the FRW  metric (Eq. 24 from \cite{Carroll1992}) and introduce $R_0$ on both sides of the relationship
\bea
\frac{d R_0r}{dt}&=&\frac{R_0}{R}(1-kr^2)^{1/2} \nnc
R_0\frac{d r}{dt}&=&\frac{1}{a(t)}(1-kr^2)^{1/2} 
\eea

\be
R_0\frac{dr}{(1-kr^2)^{1/2} }= \frac{d t}{a(t)}
\label{nullgeod}
\ee
and substitute the differential using Eq.(\ref{dt})
\be
R_0\frac{dr}{(1-kr^2)^{1/2} }=\frac{1}{a^2} \frac{da}{\displaystyle \sqrt{\frac{8\pi G}{3}\rho_{m/r}  + \frac{\Lambda(t) }{3} -\frac{k}{R_0^2a^2}}}. \nn
\ee

\nd We use the typical definition of critical mass density as
\be
 \rho_c= \frac{3H_{0}^2}{8\pi G}
\ee
and introduce the usual normalized parameters as
\be
\Omega_i=\frac{\rho_{i,0}}{\rho_c} ; \quad \Omega_k=-\frac{k}{R_0^2 H_0^2}; \quad\Omega_\Lambda=\frac{\Lambda(t_0)}{3 H_0^2} 
\label{Omdef}
\ee

\nd with the sum of the various $\Omega$ terms normalized to 1 as presented as Eq.(\ref{norm}). The general $\Omega_i$ term represents the various forms of normalized matter and energy except DE.

\par We continue the derivation using Eqs.(\ref{Lambda3}, \ref{dt} and \ref{Omdef}).
\bea
\frac{R_0 dr}{\sqrt{1+\Omega_k R_0^2 H_0^2 r^2}} =\frac{1}{a^2}\frac{da}{\displaystyle \sqrt{\frac{8\pi G}{3}\rho_{m/r}  + \frac{\Lambda(t) }{3} -\frac{k}{R_0^2a^2}}}. \nnc
\frac{R_0 dr}{\sqrt{1+\Omega_k R_0^2 H_0^2 r^2}} =\frac{1}{a^2}\frac{ da}{\displaystyle H_0 \sqrt{\frac{8\pi G}{3H_0^2}\rho_{m/r}  + \frac{\Omega_\Lambda}{a^3}  + \frac{\Omega_k}{a^2}}}
\label{DL0}
\eea

\be
\int_0^r\frac{H_0R_0 dr}{\sqrt{1+\Omega_k R_0^2 H_0^2 r^2}}=\int_0^a\frac{1}{a^2}\frac{ da}{\displaystyle  \sqrt{\frac{8\pi G}{3H_0^2}\rho_{m/r}  + \frac{\Omega_\Lambda}{a^3}  + \frac{\Omega_k}{a^2}}} \nn
\ee
\nd substituting with the variable $y$ on the left hand side $\sqrt{\Omega_k}R_0H_0r=y$
the integral of the left hand side can be solved as $\int\frac{dy}{\sqrt{1+y^2}}=\ash(y)$ which leads to the relationship
\be
\frac{1}{\sqrt{\Omega_k}}\ash(\sqrt{\Omega_k}R_0H_0r)=\int_0^a\frac{1}{a^2}\frac{ da}{\displaystyle  \sqrt{\frac{8\pi G}{3H_0^2}\rho_{m/r}  + \frac{\Omega_\Lambda}{a^3}  + \frac{\Omega_k}{a^2}}}\nn
\ee

\nd and substituting to allow a measurable $D_{L}$ using the relationships \\ $D_{M} = R_{0}r$ and $D_{L} = D_{M}/a $ we arrive at 
\be D_{L}= \frac{c}{H_{0}a\sqrt{\left| \Omega _{k}\right| }}\sn \Bigg(\sqrt{\left| \Omega _{k}\right| }\int_0^a\frac{1}{a^2}\frac{ da}{\displaystyle  \sqrt{\frac{8\pi G}{3H_0^2}\rho_{m/r}  + \frac{\Omega_\Lambda}{a^3}  + \frac{\Omega_k}{a^2}}}\Bigg )
\label{genDL}
\ee
\nd where  sinn is either $sinh$ or $sin$ depending on either positive of negative spacetime curvature.

The effects of two epochs on the photon redshift will be different so we separate the descriptions of the traveling photon. The integral can be separated into two parts to handle these two situations. The first part describes the situation of photons emitted just after singularity, $t=0$, passing through the radiation dominated epoch and the second part describing the effects of traveling through the current matter dominated epoch. For the matter dominated epoch we include the effects of dust baryons and dark matter in our model as the second term in
\bea 
D_{L}&=&\frac{D_H}{a\sqrt{\left| \Omega _{k}\right| }}\sn \left ( \sqrt{\left| \Omega _{k}\right| } \left[ \int_0^{a_1}\frac{1}{a^2}\frac{ da}{\displaystyle  \sqrt{\frac{8\pi G}{3H_0^2}\rho_r + \frac{\Omega_\Lambda}{a^3}  + \frac{\Omega_k}{a^2}}}\right.  \right. \nnc &&  \hspace*{3.5cm}+
\left.   \left. \int_{a_1}^a\frac{1}{a^2}\frac{ da}{\displaystyle  \sqrt{\frac{8\pi G}{3H_0^2}\rho_m + \frac{\Omega_\Lambda}{a^3}  + \frac{\Omega_k}{a^2}}} \right) \right]
\label{genDLsep}
\eea
with $D_H=\frac{c}{H_{0}}$ (where c is light speed in km/s and $H_0$ the Hubble constant in km s$^{-1}$Mpc$^{-1}$) is the distance between the emitter and the observer. Although true there was abundant matter present during the radiation dominated epoch we shall not evaluate this situation further. 

We now separate the two portions from within the entire distance, $[0,a]$ through which light has traveled. The first portion is within the early epoch of radiation domination $[0,a_1(t_1)]$ and for this we use Eq.(\ref{rhotintrad}) for the density in Eq.(\ref{genDL}). For the second portion $[a_1(t_1),a(t)]$  we use Eqs.(\ref{rhotintrad},\ref{rhotintmat}) for the density in Eq.(\ref{genDLsep}) because the matter dominated epoch still retains some radiation.

The relationship describing the general situation during both the radiation dominated and matter dominated epochs is

\bea D_{L}&=&\frac{c}{H_{0}a\sqrt{\left| \Omega _{k}\right| }} \sn \Bigg(\sqrt{\left| \Omega _{k}\right|}\Bigg[ \int_0^{a_1}\frac{1}{a^2} \frac{da}{\sqrt{\displaystyle\frac{\Omega_r}{a^4}+\left(\frac{4}{a^3}-\frac{3}{a^4}\right)\Omega_\Lambda+ \frac{\Omega_k}{a^2}}} \\ \nn
&&\hspace*{3cm}+\int_{a_1}^a\frac{1}{a^2} \frac{da}{\sqrt{\displaystyle \frac{\Omega_m} {a^3} +\left(\frac{1}{a^3}+\frac{3\ln(a)}{a^3}\right)\Omega_\Lambda+\frac{\Omega_k}{a^2}}}\Bigg]\Bigg).
\label{combination}
\eea

\par For our current epoch we drop the first integral to simplify the relationship as

\bea D_{L}&=&\frac{c}{H_{0}a\sqrt{\left| \Omega _{k}\right| }} \sn \Bigg(\sqrt{\left| \Omega _{k}\right|}\hspace*{1mm}[f_{(a)}]\Bigg )
\label{direct.SNe}
\eea

\nd where $f_{(a)}$ is the integral

\bea f_{(a)} =\int_{a_1}^a\frac{1}{a^2} \frac{da}{\sqrt{\displaystyle \frac{\Omega_m} {a^3} +\left(\frac{1}{a^3}+\frac{3\ln(a)}{a^3}\right)\Omega_\Lambda+\frac{\Omega_k}{a^2}}}.
\label{combo.SNe.1}
\eea

\nd We allow $t_1$ to represent the time at the end of the radiation dominated epoch, concomitantly the beginning of the mass dominated epoch. This means that $a_1=a(t_1)$ is the expansion parameter at that instant, recombination. These relationships might be used to evaluate SNe Ia data for the cosmological parameters by transposing the observed redshifts to emitted frequencies, the normalized frequency drop from distant emissions being the expansion factor.

If we allow a Universe without the CC the above relationship reduces to a straightforward expression as

\begin{equation}  
D_{L}=\frac{c}{H_{0} a\sqrt{\left| \Omega _{k}\right| }} \Bigg(\sqrt{\left| \Omega _{k}\right|}
\int_{a_1}^a[\frac{1}{a^2} \frac{da}{\sqrt{ \frac{\Omega_m} {a^3} +\frac{\Omega_k}{a^2}}}] \Bigg).
\label{simpleST.SNe}
\end{equation}

\nd The different energy dependencies of matter and spacetime are easily understood as the separate terms of the denominator within the integrand. We will see that a version of the FRW generalization, the $\Omega_{k}ST$ model, can sometimes fit the SNe Ia data fairly well.

\begin{equation}  
D_{L}=\frac{c}{H_{0} a} \Bigg(
\int_{a_1}^a[\frac{1}{a} \frac{da}{\sqrt{ \frac{\Omega_m} {a} +\Omega_{\Lambda}}{a^2}}] \Bigg).
\label{FlatDE.SNe}
\end{equation}

It is important to question of the existence of a positive root for the denominators of the integral in Eq.(\ref{combo.SNe.1}). While imaginary roots may even have physical significance those solutions may not be amenable to use with astronomical observations. For instance, reality demands a real Hubble constant, $H^2\geqslant 0$. This suggests the root of the right side of a more generalized situation as in Eq.(\ref{frm1}) below, may be the inequality

$$\frac{\Omega_m} {a^3} +\left(\frac{1}{a^3}+\frac{3\ln(a)}{a^3}\right)\Omega_\Lambda+\frac{\Omega_k}{a^2}\geq0$$

\nd with all the caveats of possible spacetime curvature and an attractive rather than repulsive DE, as a negative CC. We would like to emphasize that as we look at the integrand about the current value of the expansion parameter $a$, the contribution from the CC drops considerably. Perhaps the reason why we cannot measure this effect in the laboratory is the fact that as the expansion factor, $a$, approaches $1$ today, the influence of the CC declines towards zero?

We now cast Eq.(\ref{combination}) into the terms of observable redshift with $a=1/(1+z)$ and the fact that $\int \frac{da}{a^2}\cdots=\int dz \cdots$ resulting in a familiar format as

\bea D_{L}&=&\frac{D_H(1+z)}{\sqrt{\left| \Omega _{k}\right| }} \sn \Bigg(\sqrt{\left| \Omega _{k}\right|}\Bigg[ \int_{z_1}^{\infty} \frac{dz}{(1+z)\sqrt{\displaystyle\Omega_r(1+z)^2+\Omega_\Lambda(1-3z)(1+z)+ \Omega_k}}  \nnc
&+&\int_{z}^{z_1} \frac{dz}{(1+z)\sqrt{\displaystyle \Omega_m(1+z) +(1-3\ln(1+z))\Omega_\Lambda(1+z)+\Omega_k}}\Bigg]\Bigg ).\nnc
\label{partDLz}
\eea

\nd We again simplify the above equation to the portion which might be used to determine the cosmological parameters as 

\bea D_{L}&=&\frac{D_H(1+z)}{\sqrt{\left| \Omega _{k}\right| }} \sn \Bigg(\sqrt{\left| \Omega _{k}\right|}\hspace*{1mm}[f_{(z)}]\Bigg )
\label{hubble}
\eea

\nd where $f_{(z)}$ is the integral

\bea f_{(z)} =
\int_{z}^{z_1} \frac{dz}{(1+z)\sqrt{\displaystyle \Omega_m(1+z) +(1-3\ln(1+z))\Omega_\Lambda(1+z)+\Omega_k}}.\nnc
\label{tradition}
\eea

\nd For the special case of a flat Universe the terms containing $\Omega _{k}$ are dropped and sin{\it n} becomes 1 leading to this relationship

\bea D_{L}&=&D_H(1+z)
\int_{z}^{z_1} \frac{dz}{(1+z)\sqrt{\Omega_m(1+z) +(1-3\ln(1+z))\Omega_\Lambda(1+z)}}.\nnc
\label{tradition}
\eea
\nd Note the term containing the natural logarithm must be treated only as a positive number for evaluation with observations.

The above relationships are a version of our $\Omega_{r\Lambda k}$ model that might be evaluated with astronomical data and may be considered a broad FRW generalization allowing for the phase change at recombination. Our relationships also deal more properly with the influence of radiation in our current epoch of matter domination rather than first presuming this a minor player within the matter parameter. 

\subsection{Hubble parameter dependency of the cosmological constant}
Here we assume the cosmological constant depends on the Hubble parameter and can be approximated by the relationship

\be\Lambda(t)=3A_iH^2=3A_i\frac{\dot a  ^2}{a^2}\label{CC}\ee

\nd and we place this relation in Eq.(\ref{EoS1}) by substituting with $\kappa=k/R_0^2$
\be
\frac{\dot {a} ^{2}}{a^{2}} = \frac{8\pi G}{3}\rho  + A\frac{\dot {a} ^{2}}{a^{2}} - \frac{\kappa}{a^{2}}.
\label{frm1}
\ee
\nd We now collect the $\frac{\dot {a}^2}{a^2}$ terms as
\be
(1-A)\frac{\dot {a}^2}{a^2} = \frac{8\pi G}{3}\rho   - \frac{\kappa}{a^{2}}.
\label{frm1-a}
\ee
\nd Referring to Eq. (\ref{rho_t}) we introduce the relationship for the equation of state as $p=3\rho$ with time dependency of the cosmological constant from Eq.(\ref{CC}) and arrive at these relationships 
\be
\frac{\dot {a}^2}{a^2} =\frac{1}{(1-A)}\left(\frac{8\pi G}{3}\rho   - \frac{\kappa}{a^{2}}\right)
\label{final-frm1}
\ee
\be\frac{\ddot {a}}{a} = -\frac{4\pi G}{3}\rho( 1 + 3w ) + \frac{A}{1-A}\left( \frac{8\pi G}{3}\rho-\frac{\kappa}{a^{2}}\right).
\label{final-frm2}
\ee

\nd Then multiplying by $a^2$, taking the derivative of Eq. (\ref{final-frm1}) then dividing by $2 a \dot a$ we get
\bea
2\dot a \ddot a & = &\frac{1}{1-A×}\frac{8\pi G}{3}(\dot \rho a^2 +\rho 2a \dot a) \nnc
\frac{\ddot {a}}{a}&=&\frac{1}{1-A×}\frac{8\pi G}{3}(\dot \rho \frac{a}{2\dot a} +\rho).
\label{frm1-b}
\eea

We notice that both right hand sides of Eq. (\ref{final-frm2}) and Eq.  (\ref{frm1-b}) are equal so we can write the following relationships

\bea
&& -\frac{4\pi G}{3}\rho( 1 + 3w_i ) + \frac{A}{1-A}\left( \frac{8\pi G}{3}\rho-\frac{\kappa}{a^{2}}\right)=\frac{1}{1-A×}\frac{8\pi G}{3}(\dot \rho \frac{a}{2\dot a} +\rho )  \nnc
&&-\frac{4\pi G}{3}\rho\left(1 + 3w_i -\frac{2A}{1-A}+\frac{2}{1-A}\right)=\frac{1}{1-A×}\frac{4\pi G}{3}\dot \rho \frac{a}{\dot a}+\frac{A}{1-A}\frac{\kappa}{a^{2}}\nnc 
&&-\frac{4\pi G}{3}\rho~3(1+w_i)=\frac{1}{1-A×}\frac{4\pi G}{3}\dot \rho \frac{a}{\dot a}+\frac{A}{1-A}\frac{\kappa}{a^{2}} \nnc
&&-4\pi G(1+w_i)\rho=\frac{1}{1-A×}\frac{4\pi G}{3}\dot \rho \frac{a}{\dot a}+\frac{A}{1-A}\frac{\kappa}{a^{2}}
\eea
\nd where $i=\textrm{r,m}$ for radiation or matter. We solve this situation for the density $\rho_i$ as 
\be
\rho_i=C~a^{-3(1+w)(1-A)}+\frac{3}{4\pi G}\frac{A\kappa}{\left(1-3(1+w)(1-A)\right)a^2}.
\label{integconstant}
\ee

We must evaluate the constant of integration, $C$, using parameter values at $a(t_0)=1$ and the above equation becomes

\bea
\rho_0&=&C+\frac{3}{4\pi G}\frac{A\kappa}{\left(1-3(1+w)(1-A)\right)}\nnc
C&=&\rho_0-\frac{3}{4\pi G}\frac{A\kappa}{\left(1-3(1+w)(1-A)\right)}.
\label{constr1}
\eea
\nd We use this solution to replace $C$ in Eq.(\ref{integconstant})and after rearrangement we have
\be
\rho_i=\frac{\rho_0}{a^{3(1+w)(1-A)}}+\frac{3}{4\pi G}\frac{A\kappa}{\left(1-3(1+w)(1-A)\right)}\left(\frac{1}{a^2}-1\right).
\label{sols_2}
\ee

We now simplify the above relationship by introducing the usual normalized parameters, but in slightly different than the typical forms as
\be
\Omega_i=\frac{\rho_{i,0}}{\rho_c} ; \quad \Omega_k=-\frac{k}{R_0^2 H_0^2}; \quad\Omega_\Lambda=\frac{\Lambda(t_0)}{3 H_0^2}.
\label{Omdef}
\ee
\nd We admit that the term $A$ is a very general element and to solve for this we use Eq.(\ref{CC}) again at $t_0$ with $\Lambda(t_0)=\Lambda_0$ and Eqs.(\ref{CC},\ref{final-frm1}) using the current parameter values
\be
\Lambda_0=3A_i\left(\frac{\dot a^2}{a^2}\right)_0= 3A_i\left(\frac{8\pi G}{3(1-A_i)}\rho_0   - \frac{\kappa}{(1-A_i)}\right) \nn
\ee
\be
\Lambda_0= 3A_i\left(\frac{8\pi G}{3(1-A_i)}\rho_0   - \frac{\kappa}{(1-A_i)}\right). \nn
\ee
\nd If we remember that $\rho_c=\frac{3H_{0}^2}{8\pi G}$ we can write above equation in terms of cosmological parameters as
\be
3H_0^2\Omega_\Lambda=3A_iH_0^2\left(\frac{\Omega_i}{(1-A_i)}+\frac{\Omega_k}{(1-A_i)}\right).
\label{constr2}
\ee

\nd We can simplify the above for $A_i$ in terms of the cosmological parameters arriving at the compact relationship
\be
A_i=\frac{\Omega_\Lambda}{×\Omega_\Lambda+\Omega_i+\Omega_k}.
\label{Acons}
\ee

\nd If we use Eq.(\ref{final-frm1}) and Eq.(\ref{sols_2}) we can reintroduce the EoS parameter now embedded in $\alpha_i$

\be
\frac{\dot {a}^2}{a^2} =\frac{1}{(1-A_i)}\left\{\frac{8\pi G}{3}\left[\frac{\rho_0}{a^{{\alpha_i}}}+\frac{3}{4\pi G}\frac{A_i\kappa}{\left(2-{\alpha_i}\right)a^2}\left(1-\frac{1}{a^{{\alpha_i}-2}}\right)\right]-\frac{\kappa}{a^{2}}\right\} \nn
\ee
\nd where ${\alpha_i}=3(1+w)(1-A_i)$.
\nd

We now insert the various definitions of the cosmological parameters from Eq.(\ref{Omdef})
 \be
\frac{\dot {a}^2}{a^2} =\frac{1}{(1-A_i)}\frac{H_0^2\Omega_i}{a^{{\alpha_i}}}-\frac{2A_i}{(1-A_i)}\frac{H_0^2\Omega_k}{\left(2-{\alpha_i}\right)a^2}\left(1-\frac{1}{a^{{\alpha_i}-2}}\right)+\frac{H_0^2\Omega_k}{(1-A_i)a^{2}}
\ee

\be
\frac{\dot {a}^2}{a^2}=\frac{1}{(1-A_i)}\frac{H_0^2\Omega_i}{a^{{\alpha_i}}}+\frac{H_0^2\Omega_k}{(1-A_i)a^2}\left(\frac{2A_i}{{\alpha_i}-2}-\frac{2A_i}{({\alpha_i}-2) a^{{\alpha_i}-2}}+1\right)
\ee

\bea
\dot a &=&a\sqrt{\frac{1}{(1-A_i)}\frac{H_0^2\Omega_i}{a^{\alpha_i}}+
\frac{H_0^2\Omega_k}{(1-A_i)a^2}\left(\frac{2A_i}{{\alpha_i}-2}-\frac{2A_i}{({\alpha_i}-2) a^{\alpha_i}}+1\right)}\nnc
\frac{da}{dt}&=&a\frac{H_0}{\sqrt{1-A_i}}\sqrt{\frac{\Omega_i}{a^{\alpha_i}}+\frac{\Omega_k}{a^2}\left(\frac{2A_i}{{\alpha_i}-2}
-\frac{2A_i}{({\alpha_i-2) a^{{\alpha_i}}}}+1\right)} \nnc
dt&=& \frac{\sqrt{1-A_i}}{H_0}\frac{da}{a\sqrt{\displaystyle \frac{\Omega_i}{a^{\alpha_i}}+\frac{\Omega_k}{a^2}\left(\frac{2A_i}{{\alpha_i}-2}
-\frac{2A_i}{({\alpha_i} -2)a^{\alpha_i}}+1\right)}}.
\label{dtHubble}
\eea

\nd again using Eq.(\ref{DL0}) as the null geodesic and Eq.(\ref{dtHubble}) above

\be
R_0\frac{dr}{(1-kr^2)^{1/2} }= \frac{\sqrt{1-A_i}}{H_0}\frac{da}{a^2\sqrt{\displaystyle \frac{\Omega_i}{a^{\alpha_i}}+\frac{\Omega_k}{a^2}\left(\frac{2A_i}{{\alpha_i}-2}
-\frac{2A_i}{({\alpha_i} -2)a^{\alpha_i}}+1\right)}}
\label{nullgeod}
\ee

\nd with similar operations as previously presented above we arrive at 
\be D_{L}= \frac{c}{H_{0}a\sqrt{\left| \Omega _{k}\right| }}\sn \left((\sqrt{\left| \Omega _{k}\right| }
\sqrt{1-A_i}\int_0^a\frac{da}{a^2\sqrt{\displaystyle \frac{\Omega_i}{a^{\alpha_i}}+\frac{\Omega_k}{a^2}\left(\frac{2A_i}{{\alpha_i}-2}
-\frac{2A_i}{({\alpha_i} -2)a^{\alpha_i}}+1\right)}}\right)
\label{genDLHubble}
\ee

\nd which is a very general equation describing an evolving universe from soon after singularity, through the radiation dominated epoch until the present. We can "pull apart" this relationship into two terms, the first describing the radiation dominated epoch and the second our current matter dominated epoch as

\bea 
D_{L}&=&\frac{D_H}{a\sqrt{\left| \Omega _{k}\right| }}\sn \left ( \sqrt{\left| \Omega _{k}\right| } \left[ \sqrt{1-A_{r}}\int_0^{a_1}\frac{da}{a^2\sqrt{\displaystyle \frac{\Omega_r}{a^{\alpha_r}}+\frac{\Omega_k}{a^2}\left(\frac{2A_r}{{\alpha_r}-2}-\frac{2A_r}{({\alpha_r} -2)a^{\alpha_r}}+1\right)}} \right.  \right. \nnc &&  \hspace*{3cm}+
\left.   \left. \sqrt{1-A_{m}}\int_{a_1}^a\frac{da}{a^2\sqrt{\displaystyle \frac{\Omega_m}{a^{\alpha_m}}+\frac{\Omega_k}{a^2}\left(\frac{2A_m}{{\alpha_m}-2}
-\frac{2A_m}{({\alpha_m} -2)a^{\alpha_m}}+1\right)}} \right) \right]. \nnc
\label{genDLsepHubble.1}
\eea

\nd For evaluation with SNe Ia data we shall drop the radiation term to leave

\be
D_{L}=\frac{D_H}{a\sqrt{\left| \Omega _{k}\right| }}\sn \left ( \sqrt{\left| \Omega _{k}\right| }   \sqrt{1-A_{m}}\int_{a_1}^1\frac{da}{a^2\sqrt{\displaystyle \frac{\Omega_m}{a^{\alpha_m}}+\frac{\Omega_k}{a^2}\left(\frac{2A_m}{{\alpha_m}-2}
-\frac{2A_m}{({\alpha_m} -2)a^{\alpha_m}}+1\right)}} \right) 
\label{genDLsepHubble.2}
\ee

\nd which can perhaps be evaluated numerically for the cosmological parameters, remembering the definitions of $A_i$ and $\alpha_i$ and that the current location of the SNe Ia observations is earth so that the upper integration limit, $a$, is 1. 

For a flat Universe, with $\Omega_k=$0, but with the equation of state parameter, we can use a reduced version of the above as

\be
D_{L}=\frac{D_H}{a}  \sqrt{1-A_{m}}\int_{a_1}^1\frac{da}{a^2\sqrt{\displaystyle \frac{\Omega_m}{a^{\alpha_m}}}} 
\label{genDLsepHubble.Flat}
\ee
\nd which may be integrated to

\be
D_{L}=\frac{D_H}{a\sqrt{\Omega_m}} \sqrt{1-A_{m}} \frac{1-a_1^{-1+\frac{\alpha_m}{2}}}{-1+\frac{\alpha_m}{2}}
\label{genDLflatHubble.1}
\ee

\nd where all parameters will be simplified due to flatness ($\Omega_k=0$) so the parameter relationships simplify as
 
$$A_m=\frac{\Omega_\varLambda}{\Omega_\varLambda+\Omega_m}$$
$$1-A_m=\frac{\Omega_m}{\Omega_\varLambda+\Omega_m}$$
\nd where the $ \alpha_m$ represents the mass at $w=0$ and remembering that $\Omega_{\Lambda}+\Omega_m=1$
$$\alpha_m=3(1+0)1-A_m)=3\frac{\Omega_m}{\Omega_\varLambda+\Omega_m}=3{\Omega_m}$$

\nd so that Eq.(\ref{genDLflatHubble.1}) reduces to

\be
D_{L}=\frac{D_H}{a\sqrt{\Omega_m}} \sqrt{1-A_{m}} \frac{1-a_1^{\frac{\alpha_m}{2}-1}}{\frac{\alpha_m}{2}-1}
\label{genDLflatHubble.2}
\ee

\nd which then reduces to a fairly simple relationship as

\be
D_{L}=\frac{D_H}{a}(\frac{1-a_1^{\frac{3\Omega_m}{2}-1}}{\frac{3\Omega_m}{2}-1}).
\label{genDLflatHubble.3}
\ee

\nd This is an unusual version of a flat Universe, but still allows the normalization condition of $1=\Omega_m + \Omega_{\Lambda}$. The value for $a_1$ might be calculated from the redshift at recombination which has been estimated at about 1080\cite{Silk2008}.

\subsection{ $\Lambda$ proportional to the matter density}
We shall take the reader through a brief derivation of some of the more important points for a CC with matter density dependence. Vishwakarma pointed out\cite{Vishwa2001} that if the CC displays a dependency on the Hubble constant, $H^2$, this also means that 

$$
\Lambda(t)\propto \rho 
$$
\nd which we can describe by a simple equation as 
\be
\Lambda(t)=A\rho.
\label{lambda-a-dep}
\ee

\nd If we place this relation into the two equations of state describing our Universe, which are the usual beginnings assuming the FRW approximations of homogeneous, isotropic matter and energy distributions, we start with
\be
\left(\frac{\dot {a}}{a}\right)^{2} = \frac{8\pi G}{3}\rho  + \frac{\Lambda(t) }{3} - \frac{k}{R_0^2a^{2}}
\label{frm1}
\ee
\be\frac{\ddot {a}}{a} = -\frac{4\pi G}{3}(\rho  + 3p ) + \frac{\Lambda(t) }{3}.
\label{frm2}
\ee
\nd We can arrange these equations such that we eliminate $p$ by using the familiar power law relation $p=w\rho$

\be
\left(\frac{\dot {a}}{a}\right)^{2} =\left(\frac{8\pi G}{3}+\frac{A}{3} \right)\rho- \frac{k}{R_0^2a^{2}}
\label{frmrho1}
\ee

\be\frac{\ddot {a}}{a} = \left(-\frac{4\pi G}{3}(1 + 3w)+ \frac{A }{3}\right)\rho.
\label{frmrho2}
\ee

\nd  We now multiply  Eq.(\ref{frm1}) by $a^2$ and take the derivative with respect to time, then divide by $2a\dot a$ to get 
\be\frac{\ddot {a}}{a} = \left(\frac{8\pi G}{3}+\frac{A}{3} \right)\rho+\left(\frac{4\pi G}{3}+\frac{A}{6}\right)\dot \rho \frac{a}{\dot {a}}.
\ee
\nd We now place this relationship into Eq. (\ref{frmrho2}) and with some rearrangement get 
\bea
\left(\frac{8\pi G}{3}+\frac{A}{3} \right)\rho+\left(\frac{4\pi G}{3}+\frac{A}{6}\right)\dot \rho \frac{a}{\dot {a}}&=& \left(-\frac{4\pi G}{3}(1 + 3w)+ \frac{A }{3}\right)\rho \nnc
4\pi G(1+w)\rho+\left(\frac{4\pi G}{3}+\frac{A}{6}\right)\dot \rho \frac{a}{\dot {a}}&=&0.
\nn
\eea
\nd After some more substitutions it can be shown to be equivalent to the classical first law of thermodynamics, if one allows for the proportionality constant $w$, as
\be
\dot \rho =-3\frac{1+w}{1+\frac{A}{8\pi G}}\rho \frac{\dot {a}}{a}.
\label{rho-a-dep}
\ee
\nd At this point we specify the constant $A$ as dependent on the present value of parameters $\Lambda$ and $\rho$ as
\be
\Lambda(t_0)=A\rho_{0,i}.
\label{lambda-a-dep-current}
\ee
\par We shall use Eq.(\ref{lambda-a-dep}) as applying to our Universe today but first present the usual normalized parameters
\be
\Omega_i=\frac{\rho_{i,0}}{\rho_c} ; \quad \Omega_k=-\frac{k}{R_0^2 H_0^2}; \quad\Omega_\Lambda=\frac{\Lambda(t_0)}{3 H_0^2}
\label{Omdef}
\ee
\nd where i represents the several forms of radiation and matter with the usual definitions for spacetime curvature ($\Omega_k$) and the cosmological constant ($\Omega_\Lambda$). With these definitions Eq. (\ref{lambda-a-dep-current}) becomes 
$$3H_0^2 \Omega_\Lambda=A\frac{3 H_0^2}{8\pi G}\Omega_i$$
which reduces to a simple relationship as
\be
\frac{A}{8 \pi G}=\frac{\Omega_\Lambda}{\Omega_i}.
\ee

Beginning with Eq.(\ref{rho-a-dep}) and making some substitutions we get

\be
\dot \rho =-\frac{3\Omega_i(1+w)}{\Omega_i+\Omega_\Lambda}\rho \frac{\dot {a}}{a}
\ee
\nd where the solution for $\rho$ in terms of the current matter density is
\be
\rho=\rho_0 a^{-\frac{3\Omega_i(1+w)}{\Omega_i+\Omega_\Lambda}} \nn.
\ee
\nd Substituting for $\rho_0$ with the Hubble constant, $\rho_0=\Omega_i\rho_c=\frac{3H_0^2}{8\pi G}\Omega_i$
\be
\rho=\frac{3H_0^2}{8\pi G}\Omega_i a^{-\frac{3\Omega_i(1+w)}{\Omega_i+\Omega_\Lambda}}.
\ee
\nd We can now solve for $\Lambda$, as $\Lambda(t)/3$ in terms of the cosmological parameters and the Hubble constant as
\bea \frac{\Lambda(t)}{3}&=&\frac{1}{3}8\pi G\frac{\Omega_\Lambda}{\Omega_i} \frac{3H_0^2}{8\pi G}\Omega_i a^{-\frac{3\Omega_i(1+w)}{\Omega_i+\Omega_\Lambda}} \nnc
&=&H_0^2\Omega_\Lambda a^{-\frac{3\Omega_i(1+w)}{\Omega_i+\Omega_\Lambda}}.
\eea
\nd We use this result to solve for $D_L$ in terms of the cosmological parameters by substitution with the very useful Eq.(\ref{frm1})
\bea
\frac{da}{dt} &=& a\sqrt{\frac{8\pi G}{3}\rho  + \frac{\Lambda(t) }{3} - \frac{k}{R_0^2a^{2}}}\nnc
dt&=&\frac{da}{a\sqrt{H_0^2\Omega_i a^{-\frac{3\Omega_i(1+w)}{\Omega_i+\Omega_\Lambda}}+H_0^2\Omega_\Lambda a^{-\frac{3\Omega_i(1+w)}{\Omega_i+\Omega_\Lambda}}+H_0^2\frac{\Omega_k}{a^2}}}\nnc
dt&=&\frac{da}{aH_0\sqrt{\left(\Omega_i+\Omega_\Lambda\right) a^{-\frac{3\Omega_i(1+w)}{\Omega_i+\Omega_\Lambda}}+\frac{\Omega_k}{a^2}}}.
\eea
\nd For the situation where $w=0$ and for a universe with matter, the power of $a$ is $-3\frac{\Omega_m}{\Omega_\Lambda+\Omega_m}$ and for radiation with $w=1/3$ this term becomes $-4\frac{\Omega_r}{\Omega_\Lambda+\Omega_r}$.
\\
\par Using similar substitutions we finally solve for $D_L$ as
\bea D_{L}&=&\frac{c}{H_{0}a\sqrt{\left| \Omega _{k}\right| }} \sn \Bigg(\sqrt{\left| \Omega _{k}\right|}\Bigg[ \int_0^{a_1}\frac{1}{a^2} \frac{da}{\sqrt{\displaystyle \left(\Omega_r+\Omega_\Lambda\right) a^{-4\frac{\Omega_r}{\Omega_\Lambda+\Omega_r}}+\frac{\Omega_k}{a^2}}} \\ \nn
&&\hspace*{3cm}+\int_{a_1}^a\frac{1}{a^2} \frac{da}{\sqrt{\displaystyle \left(\Omega_m+\Omega_\Lambda\right) a^{-3\frac{\Omega_m}{\Omega_\Lambda+\Omega_m}} +\frac{\Omega_k}{a^2}}}\Bigg]\Bigg).
\label{rho_dep_combination}
\eea
\nd and again the fact that $\int \frac{da}{a^2}\cdots=\int dz \cdots$ results in a familiar format as
 
\bea 
D_{L}&=&\frac{D_H(1+z)}{\sqrt{\left| \Omega _{k}\right| }} \sn \Bigg(\sqrt{\left| \Omega _{k}\right|}\Bigg[ \int_{z_1}^{\infty} \frac{dz}{(1+z)\sqrt{\displaystyle \left(\Omega_r+\Omega_\Lambda\right) (1+z)^{\frac{2(\Omega_\Lambda-\Omega_r)}{\Omega_\Lambda+\Omega_r}}+\Omega_k}}  \nnc
&+&\int_{z}^{z_1} \frac{dz}{(1+z)\sqrt{\displaystyle \left(\Omega_m+\Omega_\Lambda\right) a^{\frac{2\Omega_\Lambda-\Omega_m}{\Omega_\Lambda+\Omega_m}} +\Omega_k}}\Bigg]\Bigg ).\nnc
\label{partDLz}
\eea

\section{Evaluation of the Detailed FRW Generalization with Astronomical Observational Data}
\label{sec:2}
We evaluate several models including those presented above and other common models using the 580 SNe Ia data available as the Union2.1 compilation at this Internet link\cite{Union2.1}. We also add the earth's location for for a total of 581 data pairs. We utilise distances and errors only as megaparsec(Mpc) rather than the more commonly used, unitless log($D_L$)\cite{Oztas2008}. We evaluate the current {\it standard model}, $\Lambda$CDM, which combines baryonic matter, CDM and radiation into a single term, $\Omega_m$, while only separating DE as the independent term of $\Omega_{\Lambda}$ in a presumed flat Universe. For the {\it standard model} but also considering the term of $\Omega_k$ for spacetime, we designate this the $\Omega_{k\Lambda}$ model and contrast this model with the FRW solution without $\Omega_{\Lambda}$ but with the parameter for spacetime, $\Omega_k$, as the $\Omega_k$ST model. Another evaluation, the $\nu\Omega_k$ST model, is based on the hypothesis that the signals emitted by ancient SNe Ia were slightly bluer than present\cite{Oztas2008,bluelite}. This adjustment is placed as a term in linear combination with the curved spacetime model($\Omega_k$ST) so might be considered a "corrected" FRW model and shares many similarities to models considering an evolution of SNe Ia emissions.

\begin{figure}[h] % float placement: (h)ere, page (t)op, page (b)ottom, other (p)age
  \centering
  % file name: C:/Users/Mike Smith/Documents/Astrophysics/FriedmannEquations/Manuscript/Final/SendAway/UnionGraph.eps
  \includegraphics[bb=0 0 390 263,width=12cm,height=8.04cm,keepaspectratio]{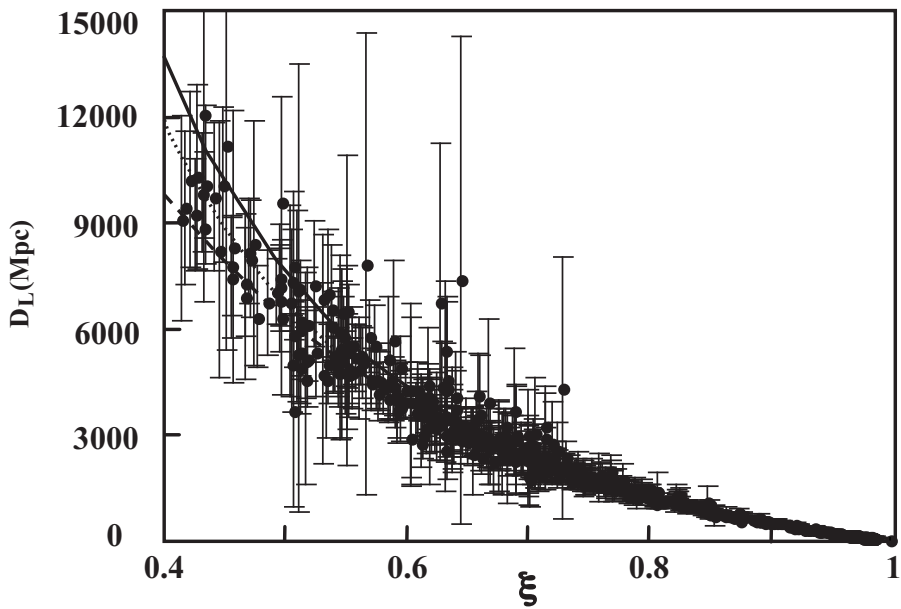}
  \caption{Best fit of SNe Ia 581 data pairs with three FRW models; distance {\it vs.} expansion factor($\xi=\nu/\nu_0$). Note the large errors associated with ancient emissions}
  \label{fig:UnionGraph}
\end{figure}

Another familiar situation, which evaluates the relationship between matter pressure and density, is dependent on the equation of state parameter $w$, which is allowed to vary during the numeric solution\cite{Carroll2012,Astier2006}. We evaluate this situation both as a flat Universe as the $\Lambda$CDM$w$ model, a variation of the {\it standard model} and as a curved Universe without the CC, the $\Omega_k$ST$w$ model. We also evaluate models which include terms for the contribution from radiation to cosmology as derived here. The results often present a very high normalized radiation parameter and never fit the data very well, so we have not included our results in any table. 

We do include evaluation of three models presented by Vishwakarma, in which the CC is dependent either on matter density, spacetime expansion or the Hubble constant, as the Vish models\cite{Vishwa2001}. These models also lump CDM, radiation and baryonic matter together and treat the spacetime contribution as implicit. In brief; his ${\it I_1}$ model is developed from the notion that the CC is inversely proportional to the Robertson-Walker metric (${\it S^-2}$), the 
 ${\it I_2}$ model where the CC is proportional to the Hubble constant (${\it H^2}$) and the (${\it I_3}$) model where the CC is proportional to the energy density. Only the variation termed ${\it I_2}$ fit the SNe Ia data well here, but we include results for all three situations. We also attempt to remedy these failures by presuming flat spacetime which greatly simplified the three models, but without an exciting results, not reported here.

We also evaluated the distance-velocity relationship proposed by Hartnett\cite{Hartnett2008} as derived from Carmeli General Relativity(CGR)\cite{Carmeli2000}. This is an interesting model in that neither spacetime nor dark energy need to be considered explicitly and it appears to model our current situation without the necessity of dark matter\cite{Hartnett2013,Oliveira}. The CGR relationships we used for modeling are

\be D_{L} = \frac{c(1+z)}{H_0\sqrt{(1-\beta^2)}}\frac{\sinh(\beta\sqrt{1-\Omega_b})}{\sqrt{1-\Omega_b}} \ee

\noindent where $\beta$ is 

\be \frac{(1+z)^2-1}{(1+z)^2+1} \ee

\noindent allowing $\Omega_b$ the normalized density for baryonic matter.

The results presented in Tables 1 through 4 are performed using calculations of the SNe Ia distances and associated errors from the modulus as $10^{((m-M)-25)/5}$ as per Riess {\it et al.}\cite{Riess1998}. It is also common to allow evaluation of the many uncertainties of these distances, $D_L$ as an extra free parameter to the modulus relationship\cite{Oliveira,Tonry2003} and these results are presented in Tables 5 through 8. The results of Tables 1,2 and 5,6 are calculated using the robust minimization routine while results in Tables 3,4 and 7,8 are calculated using the Gauss-Newton technique. The robust technique ignores outliers more than the Gauss-Newton routine; this is not a trivial difference. The robust minimization used here minimizes the sum of $Log_n(\sqrt{1+resid)^2}$, which gives preference to values close to prediction while the Gauss-Newton routine minimizes the sum of $\sqrt{resid^2}$, which does not downplay outliers. The robust routine is preferred when the data suffer significant error and has sometimes been used to analyze SNe Ia data; unfortunately the data are almost always analyzed {\it via} the coordinate transformed relative magnitude representing Log(distance) {\it vs.} logZ or Z\cite{Riess1998,Perlmutter1999}. The systematic error introduced by correlating the values in this manner does not allow confident discrimination between models.

In Table 1 we find the Vishwakarma model, ${\it I_2}$, which incorporates a $H^2$ dependent $\Lambda$, fits the data best while the $\nu\Omega_k$ST and $\Omega_{k\Lambda}$ models fit the data as well as judged by both reduced $\chi^2$ and Bayesian Information Criteria(BIC). Note that a $\Delta$BIC of less than 12 does not discriminate. These three are followed by the CRG model and other models allowing the spacetime parameter, $\Omega_k$ST and $\Omega_k$ST$w$, with lower relative values for BIC and reduced $\chi^2$ than models depending on the CC. The two {\it standard models}, which do not fit the data very well, both depending heavily on the $\Omega_{\Lambda}$ parameter. Most of these models present low values for $\Omega_m$; even the flat $\Lambda$CDM model which is by far the worst model. 

A graph of the SNe Ia data, plotted as the expansion factor, along with curves generated from the best fits of three models is shown in Figure 1. One can see the data from distant emissions needs to be carefully collected since these play a major role in model discrimination. This is a common situation where the data approaching an asymptote is important but also noisy, and the SNe Ia data are very noisy in this region.

We find the $\Lambda$CDM$w$ model fits the data best when SNe Ia distances are plotted {\it vs.} redshift using the robust routine, Table \ref{tab:2}, as judged by both BIC and reduced $\chi^2$. This fit is allowed at the expense of the equation of state parameter $w$, with a value of -0.42, which might be considered "abnormal". Surprisingly, the next best fit is the $\Omega_k$ST model, which does not consider dark energy. Neither the current {\it standard model} nor any of the Vishwakarma models are good fits as judged by both the BIC and reduced $\chi^{2}$. Note the values of the reduced $\chi^{2}$ are systematically lower than those in Table 1.  The most obvious differences between our analyses and earlier published results are the very low values for $\Omega_m$ calculated here for most models using the redshift data. This is probably due to our employment of distance errors as Mpc rather than log(Mpc).

\begin{figure}[tbp] % float placement: (h)ere, page (t)op, page (b)ottom, other (p)age
  \centering
  % file name: C:/Users/Mike Smith/Documents/Astrophysics/FriedmannEquations/Manuscript/Final/SendAway/DLversusRedshiftGraph2013.eps
  \includegraphics[bb=0 0 527 268,width=11.8cm,height=5.96cm,keepaspectratio]{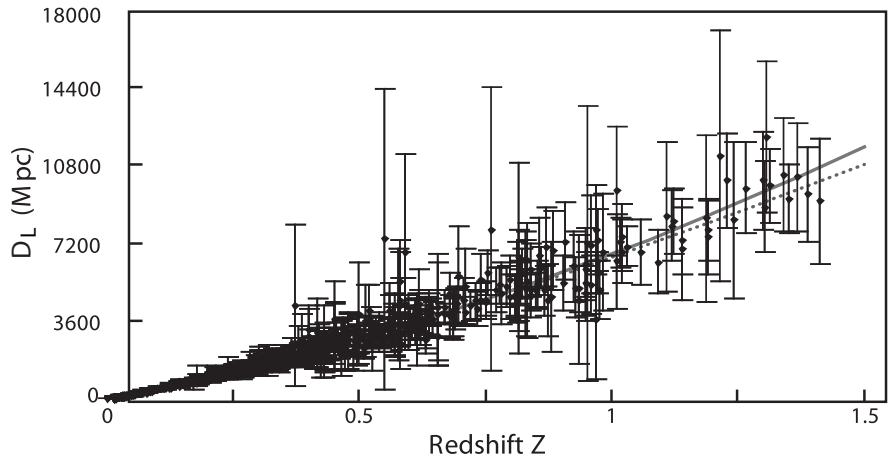}
  \caption{Best fit of SNe Ia 581 data pairs with two FRW models; distance {\it vs.} redshift, Z. Note the drastic increasing errors associated with ancient emissions compared with recent signals}
  \label{fig:DLversusRedshiftGraph2013}
\end{figure}

The SNe Ia distances as a function of associated redshifts along with curves generated from the best fits of two models are presented in Figure 2. The errors are single standard deviations; note the span of many of the observational errors are as large as the SN distances when z$>$0.5.

The opposite trends are observed when the data are analyzed using the Gauss-Newton minimization routine, as presented in Tables 3 and 4, rather than the the robust routine used to obtain the results presented in Tables 1 and 2. These two catalogues present a general preference for models based on flat spacetime and the CC. The {\it standard model} is found to fit the data best using either the expansion factor or the redshift as the abscissa. The $\Lambda$CDM$w$ and $\Omega_{k\Lambda}$ models are also pretty good fits. Another difference between the results from Tables 3,4 and 1,2 are the values for $\Omega_m$; much larger when the Gauss-Newton routine is used.

The same general ordering of goodness of fits are obtained when an additional free parameter is used during evaluations as presented in Tables 5-8. The extra parameter is sometimes termed a "nuisance" parameter\cite{Perlmutter1999} though this term is also sometimes applied to the Hubble constant (not by us). Since the SNe Ia distances admittedly suffer some degree of systematic error, in addition to a large dose of random noise especially at great distances\cite{Oliveira,Tonry2003}, adding a free parameter can partially correct for some errors. While the extra parameter barely affects the $\chi^2$ values it does increase values of $\Delta$BIC.

The results presented in Tables 5 and 6 and examination of the small differences between $\Delta$BIC values does not rule out several models as being the best fit. Table 5, with results from fits based on the expansion factor rather than redshift, suggests models including spacetime curvature are the best. The results of Table 6 suggest models including the CC fit the redshift data best.

In general, the Gauss-Newton minimization technique presents best fits for the $\Lambda$CDM model while the robust routine does not seem to have a preference. We also note that robust minimization more easily separates models both on the basis of the $\chi^2$ and BIC values.

\begin{table}
\caption{Results from least squares fits of 8 models$^a$ using a robust minimization routine of luminosity distances (Mpc) {\it vs.} the expansion factor, $\nu/\nu_0$, with $\chi^{2}/$N-FP as the reduced $\chi^{2}$ and $\Delta BIC$ as relative values}
\vskip 2 mm
\label{tab:1}       % Give a unique label
\centering
\begin{tabular}{lllllll}
\hline\noalign{\smallskip}
 Model     &   Matter Density &  & &  &   &   \\
    & $\Omega_{m}$ or $\Omega_{b}$ &$\Omega_{\Lambda}$ or $w$ & $H_0$  & $\Omega_k$ or $w$ & $\chi^{2}/$N-FP & $\Delta$BIC \\
\noalign{\smallskip}\hline\noalign{\smallskip}

$\Lambda$CDM& $0.05\pm$0.01& 0.95($\Omega_{\Lambda}$)&73.1$\pm$0.4 & - & 1.73 & 229 \\

$\Lambda$CDM$w$ & $<0.01\pm<$0.01 & $>$0.99($\Omega_{\Lambda}$)& 71.9 $\pm$0.3 & 3.91$\pm$1.12($w$) & 1.68 & 221 \\

$\Omega_k$ST$w$ & $<0.01\pm<$0.01 & 3.5$\pm>$100($w$) &68.4 $\pm$0.02 & $>0.99(\Omega_k)$ & 1.38 & 23 \\

$\Omega_k$ST & 0.02$\pm$0.04 &  -  &68.2$\pm$0.3 & 0.98($\Omega_k$) & 1.39 & 17 \\

CGR($\Omega_b$) & 0.29 $\pm$0.01 & 0.52($\Omega_{\Lambda}$) &69.9 $\pm$0.2 & 0.19$\pm$0.01($\Omega_k$) & 1.38 & 18 \\

 $\Omega_{k\Lambda}$& 0.87 $\pm$7 & 1.13($\Omega_{\Lambda}$) &72.0 $\pm$0.5 & -1$\pm$11($\Omega_k$) & 1.33 & 3 \\

$\nu\Omega_k$ST & 0.02 $\pm$0.04 & 9.7(U) &62.2 $\pm$0.3 & 0.98($\Omega_k$) & 1.33 & 3 \\

Vish$\#$2 & 0.99 $\pm$0.07 & -0.04$\pm$0.07($\Omega_{\Lambda}$) &75 $\pm$11 & -0.04($\Omega_k$) & 1.33 & 0 \\

\noalign{\smallskip}\hline
\end{tabular}
\begin{flushleft}
$^a$The $\Lambda$CDM, current {\it standard model} with flat spacetime; $\Lambda$CDM$w$, the {\it standard model} with the Equation of State parameter $w$; $\Omega_k$ST$w$, FRW model without the CC but with $w$ and curvature; $\Omega_k$ST, FRW model without the CC but with curvature; CGR, Carmeli General Relativity model with $\tau$; $\Omega_{k\Lambda}$, {\it standard model} with curvature; $\nu\Omega_k$ST, the $\Omega_k$ST model in linear combination with correction for emission blue shift.
\end{flushleft}
\end{table}

\section{Conclusions and Discussion}
Here we can broadly discern two groupings of our models with respect to ability to fit the SNe Ia data using the robust minimization routine. The first group allows spacetime curvature and while the values of $\Omega_k$ differ widely, the models fit the SNe Ia data significantly better than models without. The second groups consists of CC models allowing DE but usually do not fit the data very well. The opposite grouping is found when the minimization follows the Gauss-Newton routine; models predicting flat spacetime and the CC are preferred over those allowing curved spacetime.

Inspection of the Figures \ref{fig:UnionGraph} and \ref{fig:DLversusRedshiftGraph2013} reveals very large errors associated with distant SNe Ia events. These large uncertainties in the values for SNe Ia distances are the primary reason we cannot conclude any particular model better than the others. We doubt that if we performed analyses with the Union2.1 data for any of the multitude of models now published we could declare a best fit. While progress is being made in determining distances and reducing associated errors from several sources\cite{Kelly2010}, the current data leave much to be desired and there is more work to do to reduce the uncertainty of these measurements\cite{BH2011}. This problem is reflected by the inability of even 581 data pairs to allow discrimination between several models examined here, even though these models present drastically different physical realities. One may attempt using other statistical methods, such as "binning", but this technique has little value for analyses of these data in our hands. 

The majority of our results trend towards values for the normalization matter density which might be currently considered too high, with $\Omega_m$ over 0.30. Models without the CC most often return values for $\Omega_m$ which would be considered too low by many. Matter density much lower than proposed by the current {\it standard model} are preferred by some astronomers\cite{Kroupa2013} and are matched by BBN calculations\cite{Burles1,Burles2} which place into question the reality of dark matter\cite{Copi1995}.

We wonder about the meaning of the $k$, the constant of integration leading to Eq.\ref{EoS1}. Many think the value of $k$ only indicates the degree of spacetime curvature, with a $k$=0 indicating flat spacetime, for instance. We believe the value for $k$ means not only this but includes the relative amount of spacetime in a system. A solution of the FRW model with a small value for $\Omega_m$ and a large value for $\Omega_k$ means, as some models here present, that a universe is sparsely populated with matter separated by extremely large distances, much like our Universe.
\newpage
\section{Tabulated results}

\begin{table} [h]
\caption{Results from least squares fits of eight models$^a$ using a robust minimization routine of luminosity distances (Mpc) {\it vs.} redshifts with $\chi^{2}/$N-FP as the reduced $\chi^{2}$ and $\Delta BIC$ as relative values}
\vskip 2 mm
\label{tab:2}       % Give a unique label
\centering
\begin{tabular}{lllllll}
\hline\noalign{\smallskip}
 Model     &   Matter Density &  & &  &   &   \\
    & $\Omega_{m}$ or $\Omega_{b}$ &$\Omega_{\Lambda}$ or $w$ & $H_0$  & $\Omega_k$ or $w$& $\chi^{2}/$N-FP & $\Delta$BIC \\
\noalign{\smallskip}\hline\noalign{\smallskip}

Vish$\#$1 & 0.09$\pm$0.05 & 0.52$\pm$0.02($\Omega_{\Lambda}$) & 69.0$\pm$0.4 & -0.39($\Omega_k$) & 3.61 & 775 \\

$\Lambda$CDM & 0.001$\pm$0.01 & $>$0.99($\Omega_{\Lambda}$) & 73.1 $\pm$0.4 & - & 2.01 & 421 \\

$\Omega_k$ST$w$ & 0.02$\pm$0.006& 3.73$\pm$ 0.29($w$)&  74.0$\pm$0.5 & 0.98($\Omega_k$) &1.82 &375 \\

CGR($\Omega_{b}$) & 0.47$\pm$0.25 & - & 72.4$\pm$0.4 & - &  1.44& 231 \\

Vish$\#$2 & 0.37$\pm$0.42 & 0.80$\pm$0.13($\Omega_{\Lambda}$) & 69.4$\pm$0.5 & 0.17($\Omega_k$) & 1.40 & 212 \\

Vish$\#$3 & 0.10$\pm$2.6 & 0.66$\pm$1.2($\Omega_{\Lambda}$) & 69.3$\pm$0.4 & -0.24($\Omega_k$) & 1.38 & 202 \\

$\Omega_k$ST & 0.0002$\pm$0.03 &  - &67.7$\pm$ 1.2 & $>0.99(\Omega_k$) & 1.03 & 32 \\

$\Lambda$CDM$w$ & 0.33$\pm$0.07 & 0.67($\Omega_{\Lambda}$) &71.2 $\pm$0.3 & -0.42$\pm$0.31($w$) & 0.98 & 0 \\

\noalign{\smallskip}\hline
\end{tabular}
\begin{flushleft}
The Vish models are from reference \cite{Vishwa2001} corresponding to $\it I_{1}$, $\it I_{2}$ and $\it I_{3}$; all others as Table 1. Note the models of Vishwakarma are based on the unusual $1+\Omega_k=\Omega_m+\Omega_\Lambda$.
\end{flushleft}
\end{table}

\begin{table} [h]
\caption{Results from least squares fits of 8 common models$^a$ using a Gauss-Newton minimization routine of luminosity distances (Mpc) {\it vs.} the expansion factor, $\nu/\nu_0$ with $\chi^{2}/$N-FP as the reduced $\chi^{2}$ and $\Delta BIC$ as relative values}
\vskip 2 mm
\label{tab:3}       % Give a unique label
\centering
\begin{tabular}{lllllll}
\hline\noalign{\smallskip}
 Model     &   Matter Density &  & &  &   &   \\
    & $\Omega_{m}$ or $\Omega_{b}$ & $\Omega_{\Lambda}$ or $w$ & $H_0$  & $\Omega_k$ or $w$ &$\chi^{2}/$N-FP & $\Delta$BIC \\
\noalign{\smallskip}\hline\noalign{\smallskip}

CGR($\Omega_{b}$) & 0.54$\pm$0.2 & - & 52.5$\pm$0.2 & - & 1.36 & 33 \\

$\Omega_k$ST & $<0.001\pm$0.005 &  - &68.1$\pm$0.2 & $>0.99(\Omega_k$) & 1.37 & 38 \\

$\nu\Omega_k$ST & $0.09\pm$0.04 & - &62.1$\pm$0.3 & 0.91($\Omega_k$) & 1.33 & 23 \\

$\Omega_k$ST$w$ & 0.13$\pm$0.001 & -8.3$\pm$5.3($w$) &71.2 $\pm$1.1 &  0.87($\Omega_k$) & 1.31 & 14 \\

Vish$\#$2 & 1$\pm$0.06 & -0.02$\pm$0.06($\Omega_{\Lambda}$) &73.3 $\pm$9 & - & 1.30 & 8 \\

$\Lambda$CDM$w$ & 0.26$\pm$0.10 & 0.74($\Omega_{\Lambda}$) &70.5 $\pm$0.4 & $-0.59\pm0.22$($w$) & 1.30 & 6 \\

$\Omega_{k\Lambda}$ & 0.53$\pm$2.4 & 0.82$\pm$1.2($\Omega_{\Lambda}$)  &70.4$\pm$0.4 & -0.35($\Omega_{k}$) & 1.29 & 6 \\

$\Lambda$CDM & 0.30$\pm$0.02 & 0.70($\Omega_{\Lambda}$) & 70.4$\pm$0.3 & - & 1.29 & 0 \\

\noalign{\smallskip}\hline
\end{tabular}
\begin{flushleft}
$^a$Abbreviations as per Tables 1,2. The order of models may be different depending on values for reduced $\chi^{2}$ or $\Delta BIC$ which place different emphasis on the number of free parameters.
\end{flushleft}
\end{table}

\begin{table} [h]
\caption{Results from least squares fits of 9 models$^a$ using a Gauss-Newton minimization routine of luminosity distances (Mpc) {\it vs.} redshifts with $\chi^{2}/$N-FP as the reduced $\chi^{2}$ and $\Delta BIC$ as relative values}
\vskip 2 mm
\label{tab:4}       % Give a unique label
\centering
\begin{tabular}{lllllll}
\hline\noalign{\smallskip}
 Model     &   Matter Density &  & &  &   &   \\
    & $\Omega_{m}$ or $\Omega_{b}$ &$\Omega_{\Lambda}$ or $w$ & $H_0$  & $\Omega_k$ or $w$ & $\chi^{2}/$N-FP & $\Delta$BIC \\
\noalign{\smallskip}\hline\noalign{\smallskip}

$\Omega_k$ST & $<0.0001\pm$0.004 & - &68.1$\pm$0.3 & $>0.99(\Omega_{k})$ & 1.03 & 38 \\

CGR($\Omega_{b})$ & 0.54$\pm$0.20 & - & 68.5$\pm$0.3 & - &  1.02& 33 \\

Vish$\#$1 & 0.53$\pm$0.64 & 0.47$\pm$0.21($\Omega_{\Lambda}$) & 69.1$\pm$0.5 & 0($\Omega_k$) & 1.00 & 25 \\

Vish$\#$1Flat & 0.53$\pm$0.01 & 0.47($\Omega_{\Lambda}$) & 69.1$\pm$0.3 & - &  0.99 & 18 \\

Vish$\#$2 & 0.45$\pm$0.36 & 0.55$\pm$0.10($\Omega_{\Lambda}$) & 69.6$\pm$0.4 & 0($\Omega_k$) &  0.98&15 \\

Vish$\#$3 & 0.44$\pm$0.23 & 0.56$\pm$0.04($\Omega_{\Lambda}$) & 69.7$\pm$0.4 & 0($\Omega_k$) &  0.98& 15 \\

$\Omega_k$ST$w$ & $0.54\pm$.09 & 0.60$\pm$0.13(w) &70.7 $\pm$0.5 & 0.46($\Omega_k$) & 0.97 & 6 \\

$\Lambda$CDM$w$ & 0.31$\pm$0.04 & 0.69($\Omega_{\Lambda}$) & 70.5 $\pm$0.4 & -0.62$\pm$0.15($w$) & 0.97 & 6 \\

$\Lambda$CDM & 0.30$\pm$0.02 & 0.70($\Omega_{\Lambda}$) & 70.4$\pm$0.3 & - & 0.97 & 0 \\

\noalign{\smallskip}\hline
\end{tabular}
\begin{flushleft}
$^a$ Abbreviations as per Tables 1,2.
\end{flushleft}
\end{table}

\begin{table} [h]
\caption{Results from least squares fits of 7 models$^a$ using a robust minimization routine of luminosity distances (Mpc) {\it vs.} the expansion factor, $\nu/\nu_0$ and floating parameter, with $\chi^{2}/$N-FP as the reduced $\chi^{2}$ and $\Delta BIC$ as relative values}
\vskip 2 mm
\label{tab:5}       % Give a unique label
\centering
\begin{tabular}{lllllll}
\hline\noalign{\smallskip}
 Model     &   Matter Density &  & &  &   &   \\
& $\Omega_{m}$ or $\Omega_{b}$ &$\Omega_{\Lambda}$ or $w$ & $H_0$  & $\Omega_k$ or $w$ & $\chi^{2}/$N-FP & $\Delta$BIC \\
\noalign{\smallskip}\hline\noalign{\smallskip}

Vish$\#$3  &0.72$\pm>$100&0.28($\Omega_{\Lambda}$)&51.2$\pm>$1000 & - & 8.98 & 1252 \\

$\Lambda$CDM$w$ & 0.12$\pm$0.04 & 0.88($\Omega_{\Lambda}$)& 67.7 $\pm$0.9 & 0.58$\pm$0.27($w$) & 1.68&168 \\

$\Lambda$CDM&0.59$\pm$0.05&0.41($\Omega_{\Lambda}$)&67.3$\pm$0.7 & - & 1.54 & 94 \\

CGR($\Omega_{b}$) & $0.05\pm$0.23 & - & 51.6 $\pm0.3$ & - &  1.38 & 37 \\

$\Omega_k$ST & 0.02$\pm$0.04 &  -  &67.2$\pm$0.4 & 0.98($\Omega_k$) & 1.35 & 0 \\

$\nu\Omega_k$ST & 0.22 $\pm$0.05 &  -   &60.4 $\pm$0.4 & 0.78($\Omega_k$) & 1.34 & 0 \\

$\Omega_k$ST$w$ & 0.87$\pm$0.54 & -1.14$\pm>$0.04($w$)  &67.4$\pm$0.9 & 0.13($\Omega_k$) & 1.33 & 0 \\

\noalign{\smallskip}\hline
\end{tabular}
\begin{flushleft}
$^a$Abbreviations as per Tables 1,2.
\end{flushleft}
\end{table}

\begin{table} [h]
\caption{Results from least squares fits of 7 common models$^a$ using a robust minimization routine of luminosity distances (Mpc) {\it vs.} redshifts with floating parameter; ordered as the reduced $\chi^{2}$ and $\Delta BIC$ as relative values}
\vskip 2 mm
\label{tab:6}       % Give a unique label
\centering
\begin{tabular}{lllllll}
\hline\noalign{\smallskip}
 Model     &   Matter Density & & &  &   &   \\
    & $\Omega_{m}$ or $\Omega_{b}$ &$\Omega_{\Lambda}$ or w & $H_0$  & $\Omega_k$ or $w$& $\chi^{2}/$N-FP & $\Delta$BIC \\
\noalign{\smallskip}\hline\noalign{\smallskip}

Vish$\#$3 & 0.29$\pm$0.40 & 0.62$\pm$0.31($\Omega_{\Lambda}$) &67.6 $\pm$0.9 & -0.09($\Omega_k$) & 1.30 & 152 \\
 
Vish$\#$2 &0.38$\pm$0.49 &0.62$\pm$0.13($\Omega_{\Lambda}$) &67.5$\pm$0.9 & 0($\Omega_{k}$) &1.10 &71 \\

CGR($\Omega_{b}$) & $0.05\pm$0.23& - &  67.4$\pm$0.4 & -  & 1.07 & 30 \\

$\Omega_k$ST$w$ & 0.97$\pm$0.5 & 0.13$\pm$0.40($w$) &68.0 $\pm$1.4 & 0.03($\Omega_k$) & 1.05 & 24 \\

$\Omega_k$ST & $<0.001\pm$0.04 &  - &67.3$\pm$1.5 & $>0.99(\Omega_k)$ & 1.03 & 12 \\

$\Lambda$CDM & 0.46$\pm$0.03 &0.54($\Omega_{\Lambda}$)  &67.6 $\pm$0.6 & - & 1.03 & 8 \\

$\Lambda$CDM$w$ & 0.41$\pm$0.26 & 0.59($\Omega_{\Lambda}$) & 59.2 $\pm$12.7 & -0.98$\pm$0.11($w$) & 1.01 & 0 \\

\noalign{\smallskip}\hline
\end{tabular}
\begin{flushleft}
$^a$Abbreviations as per Tables 1,2.
\end{flushleft}
\end{table}
\begin{table} [h]
\caption{Results from least squares fits of 9 models$^a$ using a Gauss-Newton minimization routine of luminosity distances (Mpc) {\it vs.} redshifts with $\chi^{2}/$N-FP and floating parameter, as the reduced $\chi^{2}$ and $\Delta BIC$ as relative values}
\vskip 2 mm
\label{tab:7}       % Give a unique label
\centering
\begin{tabular}{lllllll}
\hline\noalign{\smallskip}
 Model     &   Matter Density &  & &  &   &   \\
 & $\Omega_{m}$ or $\Omega_{b}$ &$\Omega_{\Lambda}$ or $w$ & $H_0$  & $\Omega_k$ or $w$ & $\chi^{2}/$N-FP & $\Delta$BIC \\
\noalign{\smallskip}\hline\noalign{\smallskip}

CGR($\Omega_{b}$) & 1      &     - & 67.1$\pm$0.3 & - &0.99 & 13 \\

$\Omega_k$ST & 0.007$\pm$0.03 & - &67.0$\pm$1.5 & $>0.99(\Omega_k)$ & 0.99 & 13 \\

Vish$\#$1 & 0.55$\pm$0.79 & 0.45$\pm0.22(\Omega_{\Lambda})$ & 68.0$\pm$1.0 & 0($\Omega_k$) & 0.98&14 \\

Vish$\#$2 & 0.49$\pm$0.49 & 0.51$\pm0.12(\Omega_{\Lambda})$ & 68.7$\pm$0.9& 0 ($\Omega_k$) & 0.98&10 \\

Vish$\#$3 & 0.49$\pm$0.34 & 0.51$\pm0.05(\Omega_{\Lambda})$ & 68.7$\pm$0.9 & 0($\Omega_k$) &  0.98 & 10 \\

Vish$\#$1Flt & 0.56$\pm$0.02 & 0.44($\Omega_{\Lambda}$) & 68.0$\pm$0.5 & - & 0.98& 7 \\

$\Omega_k$ST$w$ & 0.53$\pm$0.13 & 0.61$\pm$0.17($w$) &70.8 $\pm$1.0 & 0.47($\Omega_k$) & 0.97 & 6 \\

$\Lambda$CDM$w$ & 0.31$\pm$0.04 & 0.69$(\Omega_{\Lambda})$ & 70.2 $\pm$1.1 & -0.65$\pm$0.17($w$) & 0.97 & 6 \\

$\Lambda$CDM & 0.31$\pm$0.03 & 0.69($\Omega_{\Lambda}$) & 70.1$\pm$0.3 & - & 0.97 & 0 \\

\noalign{\smallskip}\hline
\end{tabular}
\begin{flushleft}
$^a$Abbreviations as per Tables 1,2. Vish$\#$1Flt is solution $\it I_{1}$ from \cite{Vishwa2001} as a flat universe.
\end{flushleft}
\end{table}

\begin{table} [h]
\caption{Results from least squares fits of 8 models$^a$ using the Gauss-Newton minimization routine of luminosity distances (Mpc) {\it vs.} the expansion factor, $\nu/\nu_0$ and floating parameter; organized with $\chi^{2}/$N-FP as the reduced $\chi^{2}$ and $\Delta BIC$ as relative values}
\vskip 2 mm
\label{tab:8}       % Give a unique label
\centering
\begin{tabular}{lllllll}
\hline\noalign{\smallskip}
 Model     &   Matter Density &  & &  &   &   \\
& $\Omega_{m}$ or $\Omega_{b}$ &$\Omega_{\Lambda}$ or $w$ & $H_0$  & $\Omega_k$ or $w$ & $\chi^{2}/$N-FP & $\Delta$BIC \\
\noalign{\smallskip}\hline\noalign{\smallskip}

CGR($\Omega_{b}$) & 1$\pm>$100 & - & 51.0 $\pm$0.3 & - & 2.03 & 329 \\

$\Omega_{k}$ST & 0.006$\pm$0.04 & - &67.2$\pm$0.4 & $>$0.99($\Omega_k$)&1.33& 13\\

$\Omega_k$ST$w$& 0.14$\pm$0.11 & - & 71.6 $\pm$4.7 & -8.7$\pm$9($w$) & 1.32 & 14 \\

$\nu\Omega_{k}$ST & 0.17$\pm$0.05 & - & 61.0 $\pm$0.4 & 0.83($\Omega_k$) & 1.32 & 7 \\

Vish$\#$2 &0.96$\pm$0.08 & 0.02$\pm$0.08($\Omega_{\Lambda}$) & 66.7$\pm$12 & -0.02($\Omega_k$) & 1.30 & 7 \\

$\Lambda$CDM$w$ & 0.30$\pm$0.16 &  0.70($\Omega_{\Lambda}$)&70.2$\pm$0.9 & -0.65$\pm$0.27($w$) & 1.30 & 6 \\

$\Omega_k{\Lambda}$ & 0.86$\pm$3 & 0.96$\pm$1.5($\Omega_{\Lambda}$) &70.1$\pm$0.7 & -0.82($\Omega_k$) & 1.30 & 6 \\

$\Lambda$CDM & 0.31$\pm$0.03 &  0.69($\Omega_{\Lambda}$)  &70.1$\pm$0.6 & - & 1.29 & 0 \\
  
\noalign{\smallskip}\hline
\end{tabular}
\begin{flushleft}
$^a$Abbreviations as per Tables 1,2.
\end{flushleft}
\end{table}
%
%
%
%
% Set the ending of a LaTeX document
\end{document}